# Structural, electrical and magnetic properties of magnetoelectric GdMnO₃ thin films prepared by sol-gel method

Y. Romaguera-Barcelay[a], J. Agostinho Moreira[a], A. Almeida[a], P. B. Tavares[b] and J. Pérez de la Cruz[c,*]

[a] *IFIMUP and IN- Institute of Nanoscience and Nanotechnology, Departamento de Física da Faculdade de Ciências da Universidade do Porto. Rua do Campo Alegre, 687, 4169-007 Porto. Portugal.*

[b] *CQVR and Chemistry Department, University of Trás-os-Montes and Alto Douro, Apartado 1013, 5001-801 Vila Real, Portugal*

[c] *INESC Porto, Rua do Campo Alegre, 687, 4169-007 Porto. Portugal.*

*\*Corresponding author: e-mail: jcruz@inescporto.pt, Tel:+351220402301, Fax:+351220402437*

ABSTRACT

In this work, we analysed GdMnO₃ magnetoelectric thin films prepared by a general sol-gel method. The film formation temperature is analysed by termogravimetric analysis, while the lattice parameters of the *Pbnm* orthorhombic films are determined from the analysis of the x-ray spectra. The x-ray results also reveal that GdMnO₃ films are under compression along the *b* axis. The lattice dynamic analysis carried out by Raman spectroscopy confirms the formation of films with a *Pbnm* orthorhombic structure. Moreover, the analysis of the Raman spectra suggests that besides film-substrate interaction, texture and grain size influence can alters lattice dynamics relative to bulk ones. Magnetic measurements shows that the films response is different of the observed in ceramic and single crystal; mainly, in the shape and temperature where the transitions taking place. The magnetic study also reveals that GdMnO₃ films are paramagnetic above 80 K, showing a ferromagnetic response at low temperatures

(T≤15 K). The dielectric analysis carried out in the 850ºC annealed films shows the formation of a relaxation process at low temperature, which is associated with a polaronic process. Moreover, it is also observed a small anomaly at T~27 K that might be related with the magnetic transition taking place in this temperature range.

**1. Introduction**

Materials exhibiting coupling between electric and magnetic polarizable orders (magnetoelectrics) have attracted a lot of attention in the last decade, as they provide an additional degree of freedom in novel multifunctional devices [1–3]. Moreover, it has been found that in a large group of the magnetoelectric materials, the coupling between magnetic and polar interactions induces a ferroelectric state in an antiferromagnetic phase, which can be modified by applying an external magnetic field [4, 5].

Among the magnetoelectric materials, orthorhombic rare-earth manganites ($RMnO_3$ with R = Sm, Eu, Gd, Tb and Dy) are excellent materials for the study of the correlation between electric and magnetic properties (magnetoelectric effect), because they can be switched to the magnetoelectric state (Sm, Eu and Gd) or modified the magnetoelectric state (Tb and Dy) by applying an external magnetic field and/or tuning of the manganese ($Mn^{3+}$) magnetic structure, through changes of the rare-earth ionic radius that result in changes in the octahedral tilting angle [6–12].

The magnetoelectric transition in the orthorhombic structure is currently understood with the help of two important aspects: (i) magnetic frustration due to competing exchange integrals between successive neighbours that stabilizes a spiral magnetic phase below the Néel temperature ($T_N$) [13]; (ii) the spin–lattice coupling, where the oxygen atoms are pushed off the Mn–Mn bond, driving an electric polarization below $T_N$ [14]. As a result, this particular spin–lattice coupling mixes polar phonon and spin waves involving deviations out of the spiral magnetic plane [14].

Gadolinium manganite (GdMnO$_3$) belongs to the orthorhombic rare-earth manganite family that is magnetoelectric at low temperature. At room temperature, this system shows an orthorhombic distorted perovskite lattice structure with *Pnma* or *Pbnm* symmetry, as reported by Noda *et. al.,* [15] and Penã *et. al.*, [16]. Electric, magnetic and magnetoelectric properties of GdMnO$_3$ single crystal and ceramic have been studied systematically by several authors [10, 16–19]. At $T_N \approx 43$ K the first magnetic phase transition takes place, which is associated to an incommensurate sinusoidal antiferromagnetic order of the Mn$^{3+}$ magnetic moments [20, 21]. Due to the spin-lattice coupling, GdMnO$_3$ also exhibits a lattice modulation characterized by a propagation vector (0,$k_l$,0) below $T_N$ (Mn) [17]. The $k_l$ decreases with decreasing temperature from ~0.48 at $T_N$ up to 0.39 and then abruptly jumps to 0 at $T_{AFM1}$ ~23 K. This means that the long wavelength modulated structure vanishes below $T_{AFM1}$. The lattice modulation between $T_N$ and $T_{AFM1}$ is similar to the modulated spin structure that has been found in the TbMnO$_3$ and DyMnO$_3$ systems. However, the disappearance of the lattice modulation below $T_{AFM1}$ in the GdMnO$_3$ system implies that the long wavelength modulated spin structure is transformed to another structure. Based on the x-ray analysis and the magnetization study, it was determined that below $T_{AFM1}$ the incommensurate sinusoidal antiferromagnetic structure turns into an *A*-type canted antiferromagnetic structure [6, 17].

Finally, the magnetic anomaly observed below $T_N^{Gd} \approx 6.5$ K in this system has been associated to a long-range ordering of the gadolinium (Gd$^{3+}$) moments, related to the interaction of the 4f spins. It has been also reported that, under an external magnetic field, the GdMnO$_3$ system shows a ferroelectric state near to $T_{FE} \approx 12$K, which depend on the intensity and direction of the external magnetic field [6]. It is believed that at this temperature the Gd 4f-spin sublattice antiferromagnetically coupled with respect to Mn 3d spin and the weak ferromagnetism of $Mn^{3+}$, due to the Dzyaloshinskii–Moriya interaction, is suppressed [14, 15, 22].

Instead studying $GdMnO_3$ single crystals and ceramics, as is the general case [23, 24], it could be suitable to study the $GdMnO_3$ system in the thin film form. Even with their granular nature, films show a lot of interest, allowing a systematic comparison between the behaviour of single crystal and ceramic [5, 25]. Moreover, for several practical applications thin films are more attractive than single crystals and ceramics, due to their low cost and versatility, low sintering temperature, preparation simplicity, composition control, etc., and also because the thin film properties can somehow differ from bulk ceramics due to the substrate influence.

In this work, we report the processing procedure of orthorhombic $GdMnO_3$ thin films by a chemical solution method. A detailed analysis of their structure, microstructure and surface morphology is presented and the influence of the annealing temperature in these parameters is discussed. Lattice dynamic was carried out by Raman spectroscopy to check structural data and search for effects due to film-substrate clamping and films texture and grain size. Moreover, it is analysed the magnetic and dielectric response of the orthorhombic $GdMnO_3$ films annealed at 850 ºC, in the 5 K to 300 K temperature range and the results compared with those reported for film, ceramic and single crystal.

**2. Experimental procedure**

*2.1 Precursor solution and films preparation*

$GdMnO_3$ thin films were prepared by sol-gel method using the procedure reported by Romaguera *et. al.,* [26]. In order to obtain the precursor solution, gadolinium (III) acetate hydrate (99.99%) was dissolved in a 2:1 acetic/nitric acids molar mixture and afterward, a stoichiometric molar content of manganese (II) acetate tetrahydrate (99.99%) was added. The resulting solution was stabilized with pure 2-methoxyethanol until a final concentration of 0.2 molar.

GdMnO$_3$ precursor solution was deposited onto platinum metalized substrates (Pt/Ti/SiO$_2$/Si) supplied by Silicon Valley Microelectronics, Inc., using a Laurell WS-400-6NPP automatic spin-coater. Each individual layer was deposited at 3000 rpm during 60 seconds, dried for one minute at 80ºC and pre-sintering at 400ºC during 10 minutes in a tubular furnace. This cycle was repeated 9 times. Finally, the resulting pre-sintered films were annealed at 750ºC, 800ºC and 850ºC for an hour, followed by a quenching at 25ºC in air atmosphere, achieving a multilayer films with ~230 *nm.*

*2.2 Experimental characterization techniques*

Thermogravimetric (TG) and differential thermogravimetric analysis (DTA) were carried out in the GdMnO$_3$ dry solution, by a Seteram Labsys TG-DTA/DSC analyser, as reported elsewhere [26]. The measurements were performed in air atmosphere at a heating rate of 10ºC/minute from room temperature up to 1000ºC.

The structural characterization of the films was carried out using a PANalytical MRD diffractometer equipped with Cu K$\alpha$ radiation ($\lambda$=1.5418 Å) source in grazing angle mode with a step of 0.017º/10s from 20º up to 80º (2$\theta$).

Unpolarized micro-Raman spectra were obtained at room temperature in the 200-800 cm$^{-1}$ spectral range. The excitation was carried out with the 514.5 nm line of an Argon laser. The power incident on the sample was kept bellow 10mW in order to avoid sample heating. The scattered light was analysed by a T64000 Jobin-Yvon triple spectrometer operating in triple subtractive mode and equipped with a liquid-nitrogen-cooled charge coupled device. Identical conditions were maintained for all scattering measurements. The spectral slit width was ~1.5 cm$^{-1}$. The parameters of the observed Raman modes (frequency, line width, and amplitude) were obtained, using Igor software, from the best fit of a sum of damped oscillator functions according to the formula:

$$I(w,T) = (1 + n(w,T)) \sum_{i=1}^{N} A_{0j} \frac{w \Omega_{0j}^2 \Gamma_{0j}^2}{(\Omega_{0j}^2 - w^2)^2 + w^2 \Gamma_{0j}^2} \quad (1)$$

where $I(w,T)$ is the intensity of the disperse beam, $n(w,T)$ the Bose-Einstein factor, $A_{0j}$ the oscillator strength, $\Omega_{0j}$ the wave number and $\Gamma_{0j}$ the damping factor of the $j^{th}$ oscillator.

The morphology and roughness of the GdMnO$_3$ films surface were analysed by high resolution scanning electron microscope (SEM) FEI Quanta 400 FEG ESEM working at 15kV, while the film composition were analysed by an energy dispersive X-ray spectrometer EDAX Genesis X4M coupled to the SEM.

Magnetic properties were measured using a commercial superconducting quantum interference SQUID magnetometer in reciprocating sample option (RSO) mode with a sensitivity of ~10$^{-7}$ emu. After previous cooling down to 5 K under 0 Oe and 100 Oe applied magnetic field, respectively, the measurements were carried out in heating run from 5 K – 300 K using a driving magnetic field of 100 Oe. Magnetic hysteresis loops (*M(H)*) were measured at different temperatures using a maximum magnetic applied field of 50 kOe and a driving a.c. magnetic field of 100 Oe.

Disk-shape aluminium electrodes with 1mm in diameter and 100 nm thick were deposited by evaporation on the top of the GdMnO$_3$ films with the objective to carry out the dielectric characterization of the films. The measurements were carried out at several frequencies (1 kHz – 1 MHz) from 5 K up to room temperature by using a close-cycle cryostat system and a Hewlett-Packard precision LCR E4090A meter, controlled by a LabView® program.

**3. Results and discussion**

*3.1 Phase formation and structural characterization*

Thermogravimetric and differential thermal analysis of the GdMnO$_3$ dry solution are shown in Fig. 1. In agreement with those reported for others magnetoelectric orthorhombic films [26–29], the decomposition process of the GdMnO$_3$ solution show three major steps (*i*)- solvent evaporation, *ii*)- organic calcination, and *iii*)- phase formation). The solvent evaporation step takes place up to ~250ºC and it is characterized by a weight loss of ~15%, corresponding to the total solvent and constitutional water elimination. This step is followed by a ~40% weigh loss (see TG curve) and by the presence of various endothermic peaks in the 250-450ºC temperature range (see DTA curve), assigned to the decomposition of the organic materials in CO$_2$ and water and consequent formation of carbonates and oxycarbonates, as has been reported by several authors [26–29]. While the small weight loss observed between 450ºC–700ºC is associated to the partial decomposition of these species. Above 700ºC, there is a significant weight loss of ~2% of the initial dry solutions. This fall down, taking place around 750ºC is associated with the full decomposition of the carbonates and consequent formation of the GdMnO$_3$ phase [26–29].

Fig. 2 shows the grazing angle X-ray diffraction patterns of the GdMnO$_3$ films sintered at 750ºC, 800ºC and 850ºC, respectively. The X-ray diffraction pattern of the GdMnO$_3$ film sintered at 750ºC (Fig. 2a) shows the formation of small (121) and (002) diffraction peaks, revealing an incipient phase formation, while GdMnO$_3$ film sintered at 800ºC (Fig. 2b) exhibits the formation of a *Pbnm* orthorhombic crystalline phase, characterized by a (002) preferential orientation with α= 0.682 in March Dollase Model [30]. In the same direction, GdMnO$_3$ film sintered at 850ºC also show a *Pbnm* orthorhombic crystalline phase, with α= 0.660 in March Dollase Model for a (002) orientation.

Table 1 shows the lattice parameters of the GdMnO$_3$ films sintered at 800ºC and 850ºC and those previously reported for GdMnO$_3$ ceramic [31] and single crystal [32]. The lattice parameters obtained for as-prepared GdMnO$_3$ thin films are relatively smaller than and those reported for ceramic and single crystal, mainly in the *b* axis. This result is consistent with a compression of the GdMnO$_3$ orthorhombic *Pbnm* structure, somehow related to the film-substrate clamping.

Strain induced by the substrate (*f*) could be determined by [33]:

$$f = \frac{2(a_f - a_s)}{a_f + a_s} \sim \frac{(a_f - a_s)}{a_f} \tag{2}$$

where $a_f$ and $a_s$ are the lattice parameters of the film and free sample (ceramic or single crystal), respectively. It is visible that GdMnO$_3$ films are strongly compressed in the *b* axis (see Table 1) when compared with ceramic and single crystal. This compression in the *b* axis also increase at higher annealing temperature, which might be associated with the increase of the crystallite size and the change in the (002) film orientation.

Fig. 3 shows the scanning electron microscopy (SEM) plan-view images of the GdMnO$_3$ thin films and the EDS analysis of the 850ºC annealing sample. Film annealed at 750ºC exhibits a smooth surface without a visible grain boundary (Fig. 3a), typical of an amorphous or incipient crystalline phase. On the other hand, the GdMnO$_3$ films annealed at 800ºC (Fig. 3b) and 850ºC (Figure 3(c)) show a well-defined structure with an average grain size of 340 nm and 410 nm, respectively. The crystallization and grain growth evolution observed in the GdMnO$_3$ films are consistent with the TG-DTA and X-ray results, supporting the formation of a pure orthorhombic GdMnO$_3$ perovskite phase when the annealing temperature is higher than 750ºC.

The average films thickness calculated from the cross-section scanning electron microscopy analysis is ~230 nm in all the films, while the EDS analysis carried out in the films (unshown) exhibit peaks corresponding to the GdMnO$_3$, the carbon paste used in the SEM analysis, and the silicon, titanium, and platinum compounds present in the substrate (unshown). Semi-qualitative analysis of the EDS spectra reveals a 1:1 (Gd/Mn) molar ratio in agreement with the precursor solution composition.

Fig. 4 shows the room temperature unpolarized Raman spectra of the GdMnO$_3$ thin films annealed at 800ºC and 850ºC, along with the fitting procedure of the spectra by using equation (1). For GdMnO$_3$ orthorhombic thin films six main bands are observed, which become clearest in the sample annealed at 850ºC. The bands are located at 215 cm$^{-1}$, 369 cm$^{-1}$, 485 cm$^{-1}$, 508 cm$^{-1}$, 613 cm$^{-1}$ and 686 cm$^{-1}$, which have been associated with the GdMnO$_3$ structure (369-613 cm$^{-1}$) and with the interaction between the GdMnO$_3$ films and the metalized substrate (215 cm$^{-1}$ and 686 cm$^{-1}$) [34]. The fitting of the Raman spectra reveal nine other bands that are overlapped on the main bands or are weak $A_g$ and $B_{2g}$ Raman active bands [35]. Moreover, there are unknown bands that have been previously observed in compressed GdMnO$_3$ ceramic samples [19].

Table 2 shows a close look to the Raman bands of ceramic [34], single crystal [35] and GdMnO$_3$ thin films annealed at 800ºC and 850 ºC. It is clear that all bands belong to the GdMnO$_3$ structure, excluding the 215 cm$^{-1}$ and 686 cm$^{-1}$ bands that have been associated to a strong lattice-substrate interaction [25, 26]. It should be mentioned that the 215 cm$^{-1}$ band could be also associated with a $B_{2g}$ Raman active band, resulting of an O1 or O2 atomic motion, which could be visible due to the substrate clamping. Theoretical Raman analysis of the GdMnO$_3$ materials reveals that a $B_{2g}$ Raman active band could appear at ~200 cm$^{-1}$ [35], which is somehow consistent with the explanation mentioned above.

Six of the remaining bands reported for the films are in good agreement with the frequencies observed in GdMnO$_3$ ceramic, (i) the out of phase MnO$_6$ $x$ rotations (365 cm$^{-1}$), (ii) out of phase MnO$_6$ bending (iii) MnO$_6$ bending (482 cm$^{-1}$), (iv) O2 antistretching (502 cm$^{-1}$), (v) in phase O2 "scissorslike" (521 cm$^{-1}$) and (v) in plane O2 stretching (608 cm$^{-1}$) motions [34]. However, there is a shift to higher values in frequency of the Raman bands reported for GdMnO$_3$ thin films, mainly in the 850ºC annealed sample, which is consistent with a compressive effect. As we have reported early, the frequency of the Raman bands of polycrystalline thin films converge to the polycrystalline GdMnO$_3$ ceramic values as the film thickness increase, due to a decay of the substrate-film tension gradient [34]. Thus, beside the substrate-film interaction, the shift in the frequency of the Raman bands of the GdMnO$_3$ thin film annealed at 850ºC should be associated with other factors, like the increase in the film grain size and texture as the annealing temperature is increased.

There are other bands at 572 cm$^{-1}$ and 640 cm$^{-1}$ that have been not reported in the structural theoretical calculation [35]; however, it has been already observed in the Raman study of GdMnO$_3$ ceramic under external pressure [10, 19]. Thus, the presence of these bands is also consistent with a compression process, resulting for the formation of a shrink layer in the film-substrate interface and/or a film strained gradient due to the substrate clamping and also due to the increase in the film grain size and texture as the annealing temperature is increased.

*3.2 Magnetic and electric characterization of the GdMnO$_3$ thin films*

In order to analyse the magnetic properties of the as-prepared films, the magnetic moment of the 850ºC sintered films was measured. After cooling down the sample from room temperature up to 5 K under a zero field cooling (ZFC) and a field cooling (FC) of 100 Oe, respectively, the film was warming up from 5 K up to 300 K and the magnetic response measured by using a 100 Oe driving magnetic field. In this type of measurement, it is expected a substrate diamagnetic contribution, associated to a contribution of the silicon and silicon dioxide layers; however, in this particular case the overall magnetic response was positive, small and practically unchangeable with the temperature, as previously reported [26].

The magnetization of the 850ºC annealed GdMnO$_3$ film measured at ZFC and FC magnetic field is shown Fig. 5. The ZFC curve exhibits three anomalies, which take places in the regions reported for nanomaterials [37], ceramics [38] and single crystal GdMnO$_3$ [17, 18]. GdMnO$_3$ thin film annealed at 850 ºC shows an anomaly around ~42 K, which in GdMnO$_3$ ceramic and single crystal correspond to a paramagnetic/incommensurate antiferromagnetic phase transition and has been associated to the ordering of the Mn$^{3+}$ moments [18]; however, the wide shape of this anomaly has been not observed in GdMnO$_3$ ceramic and single crystal. This type of wide transition is normally associated to a diffuse magnetic ordering, resulting from compositional fluctuations or/and structural distortions due to a stress gradient. Thus, the wide transition observed in this film should be somehow associated with the compression effect that results for the substrate clamping and which is intensified by the modification of the film texture and grain size. Early works [39] evidenced that distorting the GdMnO$_3$ crystalline structure the Mn–O–Mn bond angle changes, altering the balance between the competing ferromagnetic and antiferromagnetic interactions and modifying the magnetic response of the sample.

Below this first anomaly at 42 K, the magnetic transitions observed in the film takes place in the vicinity of the transition temperatures of ceramic, single crystal and nanoparticles of GdMnO$_3$, which in these systems has been associated to a sine wave ordering of Mn$^{3+}$ moments and magnetic ordering of Gd$^{3+}$ moments, respectively [17, 37, 38]. As the temperature decrease, an anomaly appear at 17 K, which has been associated in ceramic and single crystal to an incommensurate antiferromagnetic - canted antiferromagnetic phase transition, due to a canting of the Mn moments and the polarization of the gadolinium 4f spins [37]. Finally, it is observed a magnetic anomaly at ~6.5 K that corresponds to a long-range ordering of the Gd$^{3+}$ moments associated with the interaction of the *4f* spins. It is believed that this Gd$^{3+}$ ordering result in a net moment of the canted Gd spins antiparallel to the canted Mn spins [37, 40].

The FC curve reveals that when the $GdMnO_3$ thin film is measured after cooling from the paramagnetic state under a field cooling of 100 Oe the magnetization increase and the low temperature transition disappears, which is consistent with the canting of the manganese and gadolinium moments.

Fig. 6 shows a series of magnetization loops performed on $GdMnO_3$ film recorded at different temperatures (5 K up to 250 K) after a ZFC process. At a maximum applied field of 50 kOe, a magnetic contribution of 99 emu/g and 74 emu/g are observed at T = 5 K and T = 15 K, respectively. The significant enhancement of the ferromagnetic properties at 5 K could be associated with the unique properties of the $Gd^{3+}$ ions, which show the largest effective spin moment (S=7/2) among all the $R^{3+}$ 4$f$ ions and the largest de Gennes factor (15.75) [41, 42]. It is visible that at very low temperatures antiferromagnetic and ferromagnetic responses coexist, which is confirmed by the magnetization loops observed in inset plot of Figure 6. There is an visible magnetization field dependence when the film is measured at T = 5 K, 15 K, showing a superposition of two magnetic contributions: a ferromagnetic component, characterized by an open hysteresis loop at relatively low coercive fields and an antiferromagnetic component, characterized by an almost linear variation of the magnetization versus the applied magnetic field at relatively high magnetic fields. On the other hand, at 34 K the magnetization curves are mainly composed by antiferromagnetic component. Above 70 K, the $GdMnO_3$ film magnetic response evolves to a linear magnetization curve, consistent with a paramagnetic state. All the hysteresis loop temperature behavior is consistent with the magnetic temperature curves observed in Figure 5, reporting an increase of the magnetic response with decreasing temperature.

Fig.s 7a) and 7b) show the temperature dependence real ($\varepsilon'$) and imaginary ($\varepsilon''$) parts of the dielectric permittivity of the 850ºC annealed $GdMnO_3$ film. There is visible higher temperature frequency dependence anomaly in the real part of the dielectric permittivity,

which is consistent with the frequency dependence peak in the imaginary part of the dielectric permittivity. Although, it has been observed a low temperature frequency dependence in the dielectric response of the GdMnO$_3$ ceramics [10], it has been not ever observed at such higher temperatures in GdMnO$_3$ single crystals [15, 17, 43], ceramics [10]. This high temperature dielectric response is associated with the injection of electric charge in the electrode film interface.

The activation energy (*Ea*) calculated from the linear fitting of relaxation time obtaining from the maximum of the ε″(T) as a function of the inverse of the temperature is shown in Fig. 7c). It reveals an *Ea$_1$*= 0.08 eV and *Ea$_2$*= 0.17 eV consistent with a polaroniclike relaxation [44]. In this type of relaxation, the dielectric loss anomaly is attributed to an interaction between free charges and the lattice which induce a local dipole [44, 45]. It is likely possible that the free charges that contribute to a polaronic relaxation with an *Ea$_1$*= 0.08 eV results from oxygen vacancies formed during film deposition process, while the polaronic relaxation with an *Ea$_2$*= 0.17 eV could result from the charge injection during the measurement process.

On the other hand, Fig. 7d) shows an enlarge of the real part of the permittivity in the temperature range of 5 K to 100 K. With decreasing temperature, a small anomaly in the permittivity is observed at T~27 K, near to the temperature where the transition from the incommensurate sinusoidal antiferromagnetic ordering of Mn$^{3+}$ to the canted structure takes place [40]. This behavior is consistent with the reported by Goto *et. al.*, [43] and Kimura *et. al.*, [17] with report a small dielectric anomaly at ~23 K for GdMnO$_3$ single crystal. This small dielectric anomaly, that take places near to the incommensurate AFM/canted AFM transition, reveals that at this temperature is quite possible that the electric response can be influenced by the magnetic response.

## 4. Conclusions

It could be concluded that above 750ºC GdMnO$_3$ thin films prepared by sol-gel methods crystallized in a *Pbnm* orthorhombic structure. GdMnO$_3$ thin films annealed at 800ºC and 850ºC show a lightly (002) preferential orientation with and crystallite size of 29.6 nm and 33 nm, respectively. In general, the lattice parameters of the films are in the order of those reported for GdMnO$_3$ ceramic and single crystal, except the *b* axis that is compressed due to the substrate clamping. The dynamic study of the films also reveals that the films crystallized in a *Pbnm* orthorhombic structure. However, there are bands attributed to the substrate-film interaction which suggest that the film structure is compressed due to the substrate clamping.

The morphology analysis shows an improvement in the film crystallization effect as the annealing temperature increase, while the stoichiometric of the film remains constant. The magnetic behavior is similar to the reported in single crystal and ceramic. Nevertheless, the shape of the anomalies and the magnetic behaviour at low temperature under an applied magnetic field reveal that the magnetic response of the films is highly influenced by the substrate. The dielectric response of the films show a relaxation process at low temperature, never observed in GdMnO$_3$ single crystal and ceramic, with is consistent with a polaroniclike relaxation. Moreover, it is observed a small anomaly at T~27 K that might be related with the magnetic transition taking place in this temperature range.

**Acknowledgements**

Authors thank Fundação para a Ciência e Tecnologia (FCT) for the financial support (PTDC/CTM/099415/2008 and PEst-C/QUI/UI0616/2011). Y. Romaguera also thanks the financial support by the Programme Alβan (The European Union Programme of High Level Scholarships for Latin America, scholarship no. E07D401169CU).

**TABLES CAPTION**

**Table 1.** Lattice parameters of the as-prepared $GdMnO_3$ thin films and reported $GdMnO_3$ ceramic and single crystal with a *Pbnm* orthorhombic structure. Moreover, the films crystallite size values and the strain induced in the $GdMnO_3$ thin films (*f*) when compared with the ceramic[a] and single crystal[b] samples.

**Table 2.** Raman bands of $GdMnO_3$ thin films annealed at 800ºC and 850ºC and the values reported for ceramic and single crystal.

**Figure Captions**

**Fig. 1**. Thermal decomposition curves of the dry $GdMnO_3$ solutions. Inset plot show phase formation region.

**Fig. 2**. Grazing angle X-ray diffraction patterns of $GdMnO_3$ thin films annealed at: a) 750ºC, b) 800ºC and c) 850ºC.

**Fig. 3.** Scanning electron microscopy images of the $GdMnO_3$ thin films annealed at: a) 750ºC, b) 800ºC and c) 850ºC.

**Fig. 4.** Room temperature Raman spectra of $GdMnO_3$ thin films annealed at: a) 800ºC and b) 850ºC.

**Fig. 5.** Temperature dependence of the magnetic response of the $GdMnO_3$ thin film annealed at 850ºC, when measured under field cooling and zero field cooling conditions.

**Fig. 6.** Magnetic response of $GdMnO_3$ thin film annealed at 850ºC, at different temperatures, under the action of external magnetic field.

**Fig. 7.** Temperature dependence of: a) the real and b) imaginary parts of the dielectric permittivity for the $GdMnO_3$ thin film annealed at 850ºC, carried out at several frequencies, c) the activation energy of the relaxation and d) a zoom of the temperature region were the magnetic transitions take place.

**Table 1.** Lattice parameters of the as-prepared GdMnO$_3$ thin films and reported GdMnO$_3$ ceramic and single crystal with a *Pbnm* orthorhombic structure. Moreover, the films crystallite size values and the strain induced in the GdMnO$_3$ thin films (*f*) when compared with the ceramic$^{(a)}$ and single crystal$^{(b)}$ samples.

| Samples | *Parameter a (Å)* | *Parameter b (Å)* | *Parameter c (Å)* | *Crystallite Size (nm)* | $f_a$ (%) | $f_b$ (%) | $f_c$ (%) |
|---|---|---|---|---|---|---|---|
| GdMnO$_3$-850ºC film | 5.309(2) | 5.776(2) | 7.442(2) | 33.0 | $^{(a)}$-0.07 | -1.32 | 0.13 |
|  |  |  |  |  | $^{(b)}$-0.17 | -1.55 | 0.15 |
| GdMnO$_3$-800ºC film | 5.318(2) | 5.783(2) | 7.428(2) | 26.9 | $^{(a)}$0.09 | -1.20 | -0.04 |
|  |  |  |  |  | $^{(b)}$0.0 | -1.43 | -0.05 |
| GdMnO$_3$ ceramic$^a$ | 5.313(1) | 5.853(1) | 7.432(1) | -- | -- | -- | -- |
| GdMnO$_3$ single crystal$^b$ | 5.318(1) | 5.866(1) | 7.431(1) | -- | -- | -- | -- |

Lattice parameters of the GdMnO$_3$ ceramic$^a$ and single crystal$^b$ taking for references [31] and [32], respectively

Table 2. Raman bands of GdMnO$_3$ thin films annealed at 800ºC and 850ºC and the values reported for ceramic and single crystal.

| Symmetry | Main atomic motion | Ceramic[a] | Single crystal[b] | Thin film 800ºC | Thin film 850ºC |
|---|---|---|---|---|---|
| interface | -- | | | 215 | 215 |
| $A_g$ | O1 displacement in *x* direction | | 276 | 269 | 272 |
| $A_g$ | In-phase MnO$_6$, *y* rotations | | 310 | 301 | 303 |
| $B_{2g}$ | O1 displacement in *z* direction | | 329 | 339 | 339 |
| $A_g$ | Out-of-phase MnO$_6$ *x* rotations | 365 | 371 | 367 | 369 |
| $B_{2g}$ | Out-of-phase MnO$_6$ bending | 465 | 469 | 467 | 469 |
| $A_g$ | MnO$_6$ bending | 482 | 486 | 483 | 485 |
| $A_g$ | O2 antistretching | 502 | 506 | 504 | 504 |
| $B_{2g}$ | In-phase O2 "scissorslike" | 521 | 525 | 523 | 523 |
| unknown | | | | 572 | 572 |
| $B_{2g}$ | In-plane O2 stretching | 608 | 611 | 610 | 613 |
| unknown | | | | 641 | 645 |
| interface | -- | | | 686 | 686 |

Raman bands of the GdMnO$_3$ ceramic[a] and single crystal[b] taking for references [34] and [36], respectively

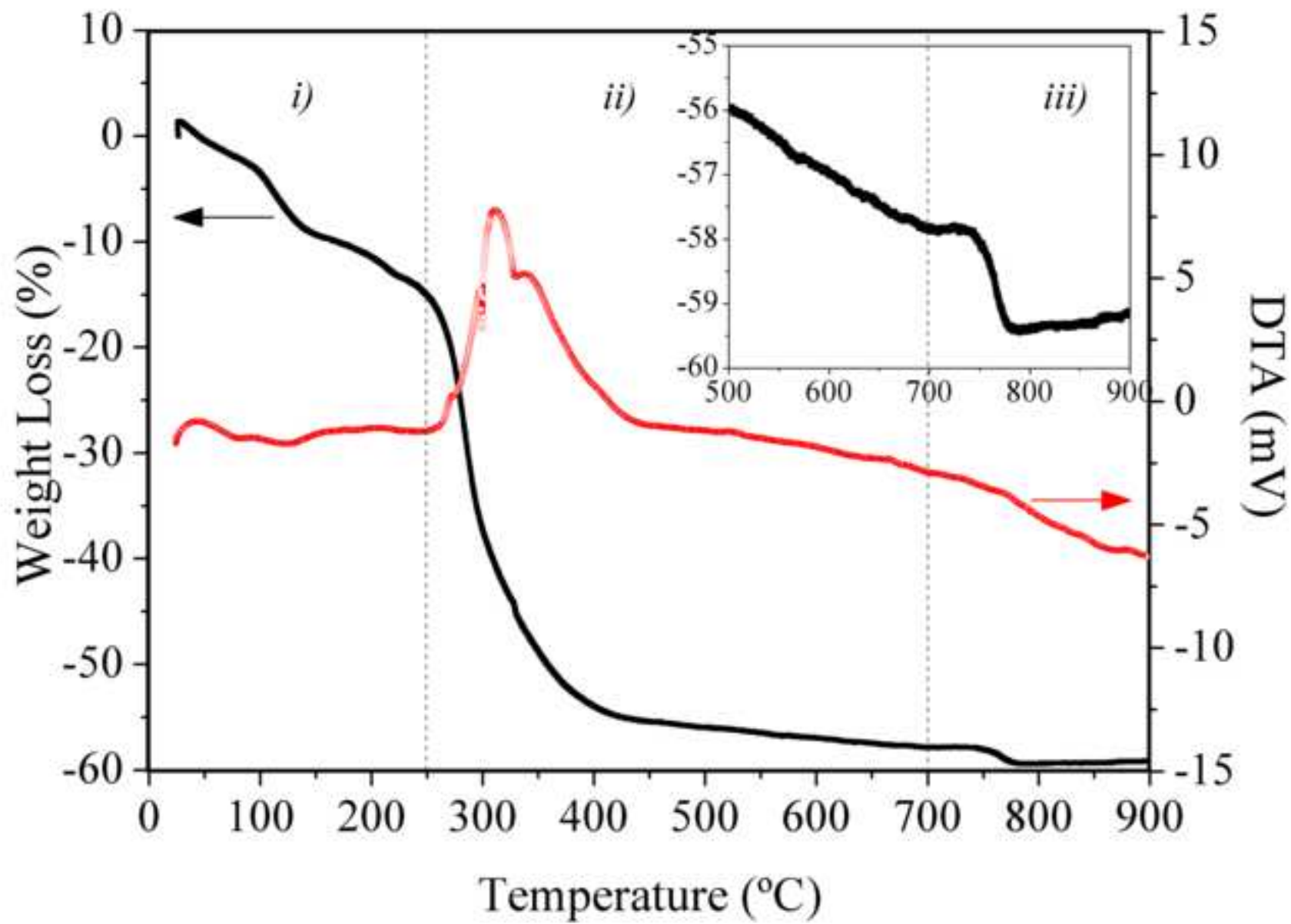

**Figure 2**
**Click here to download high resolution image**

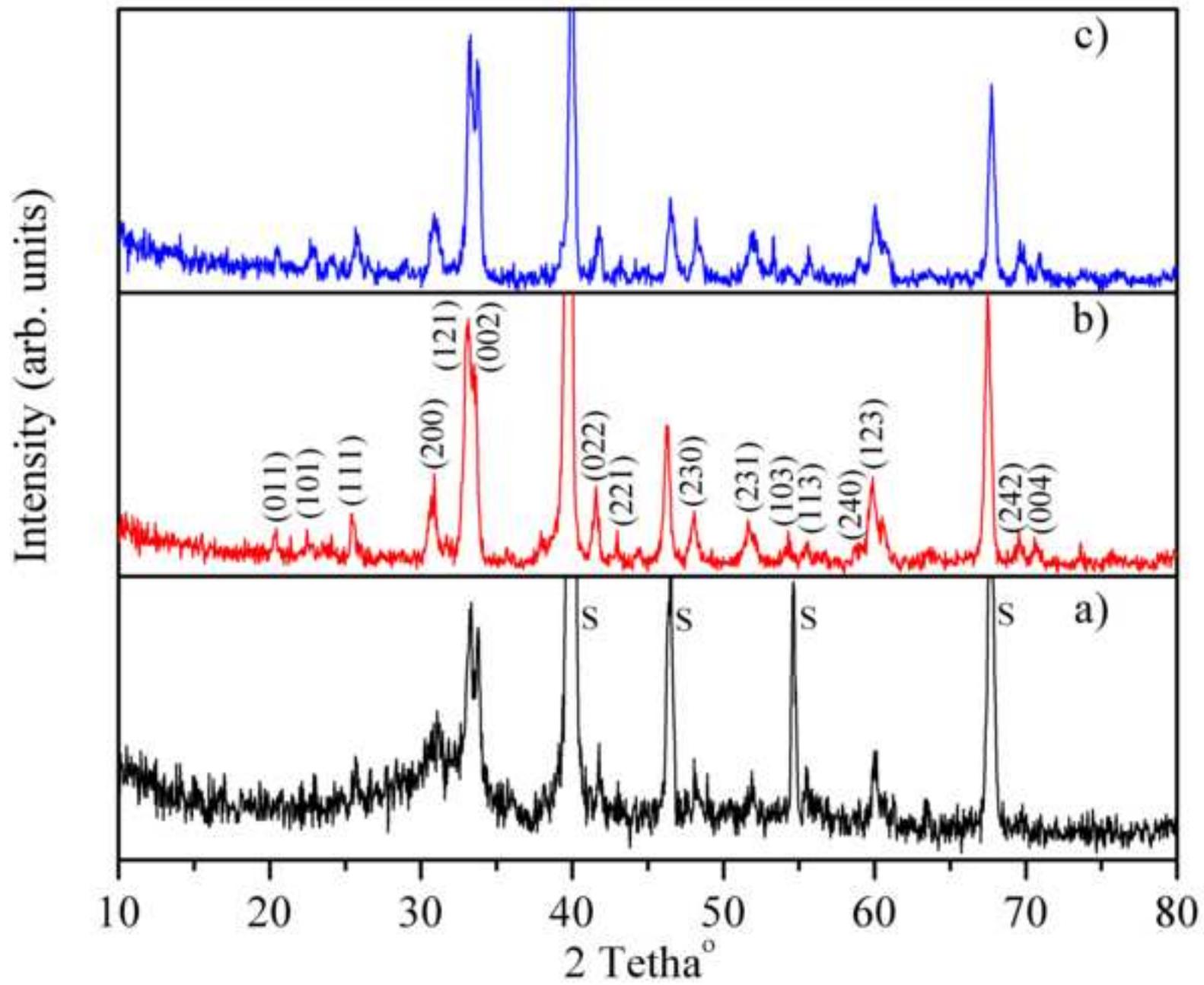

**Figure 3**
Click here to download high resolution image

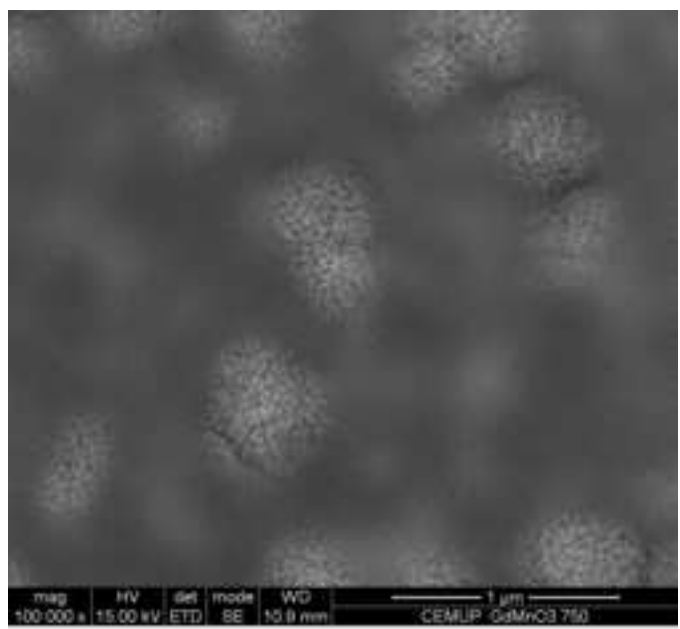
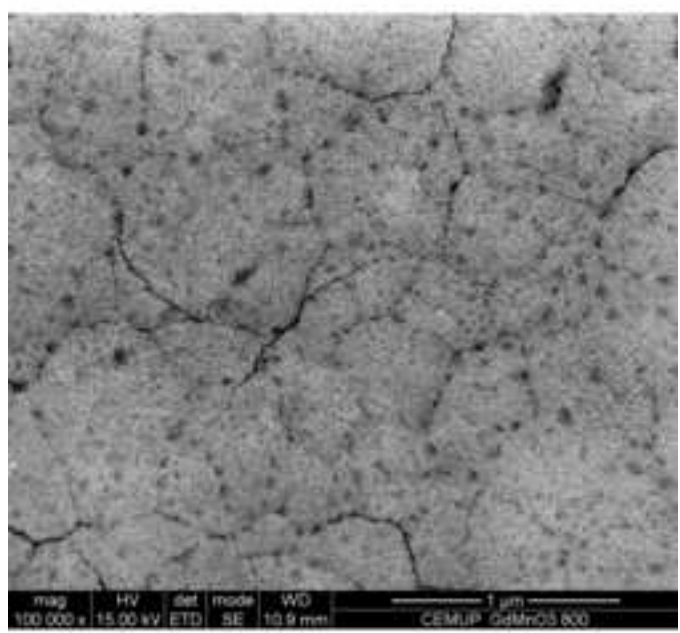
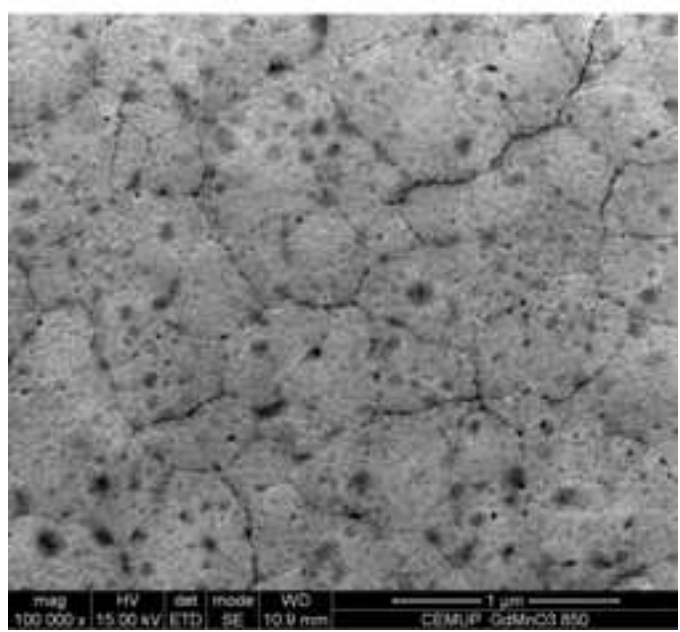

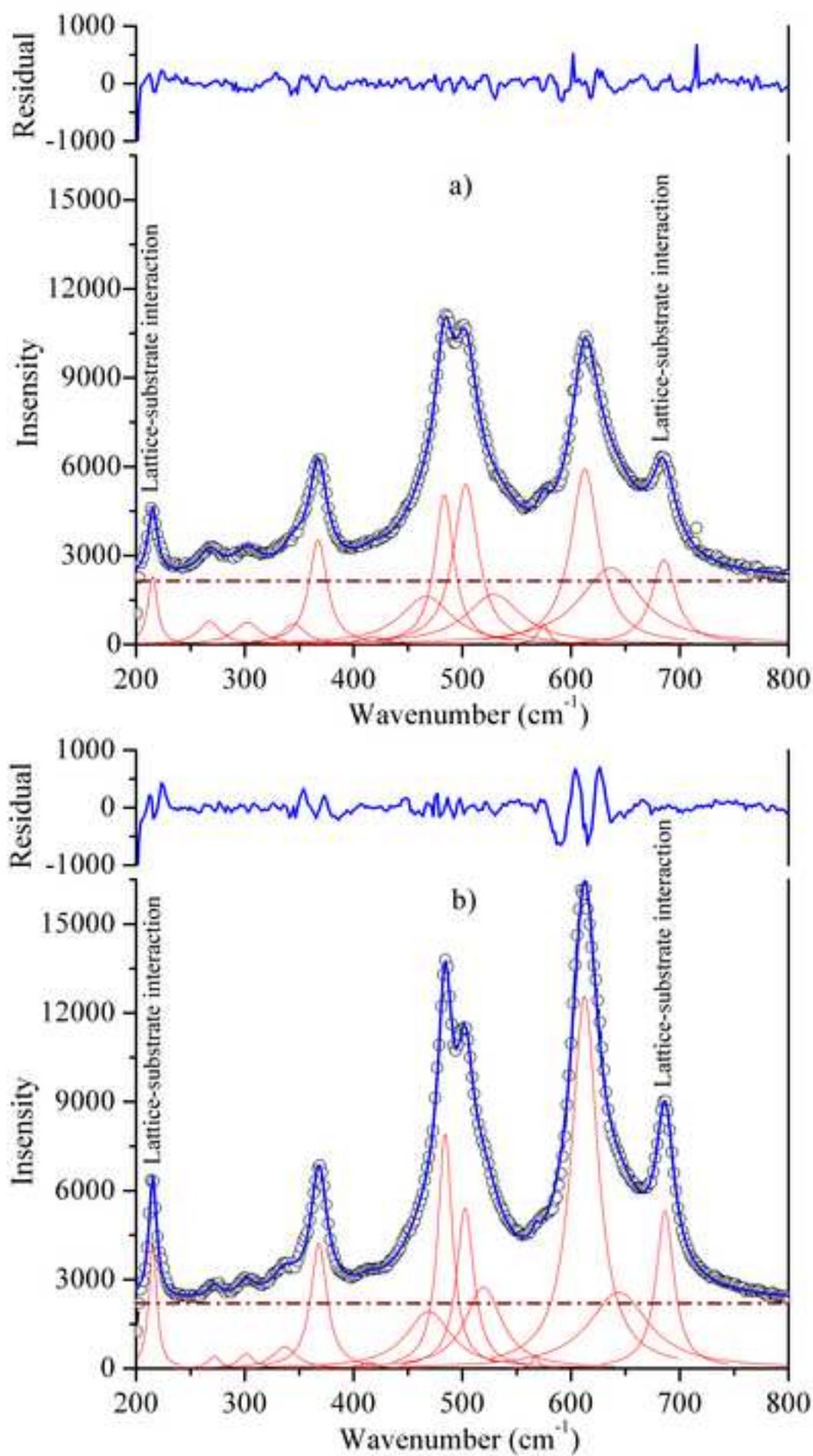

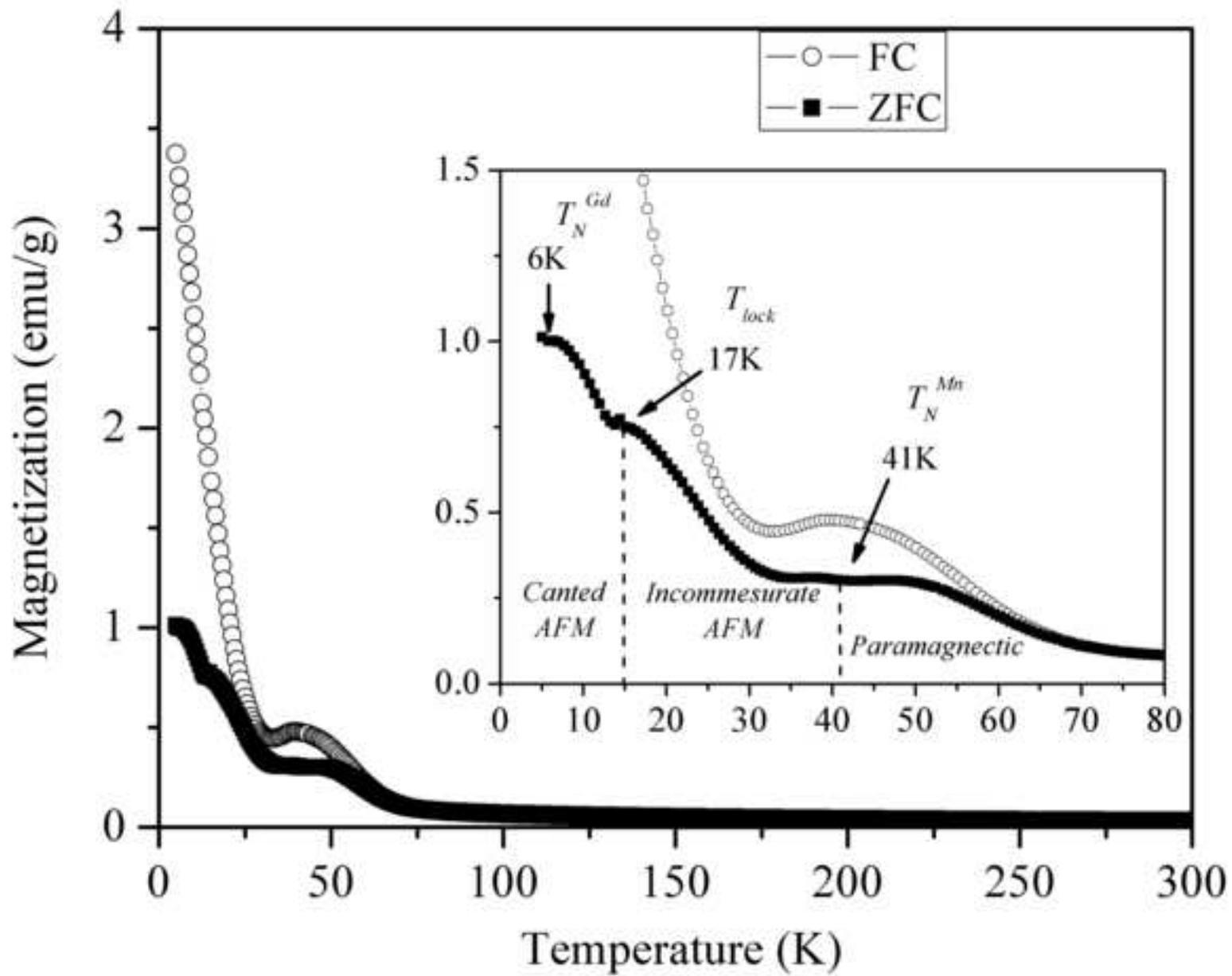

**Figure 6**
**Click here to download high resolution image**

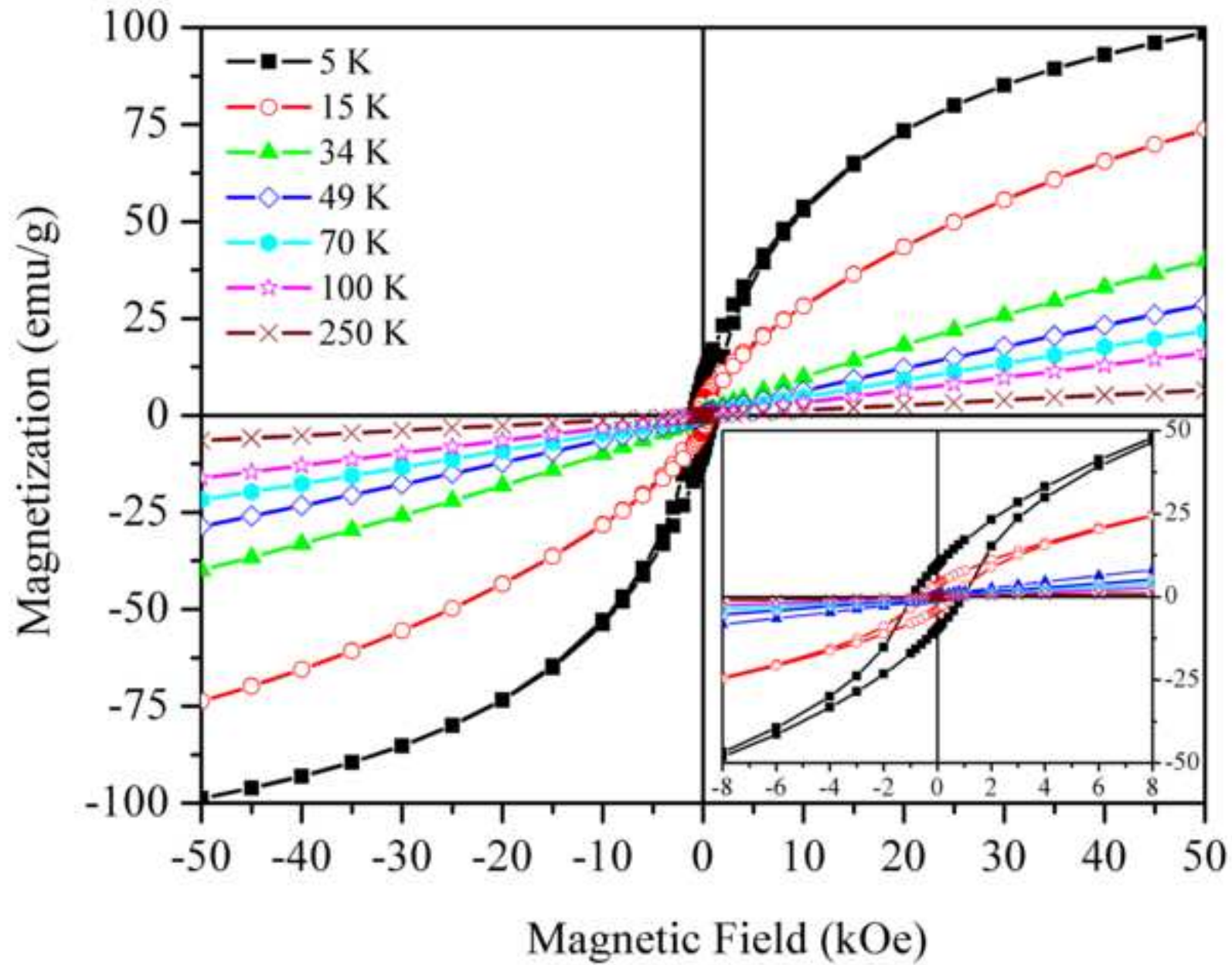

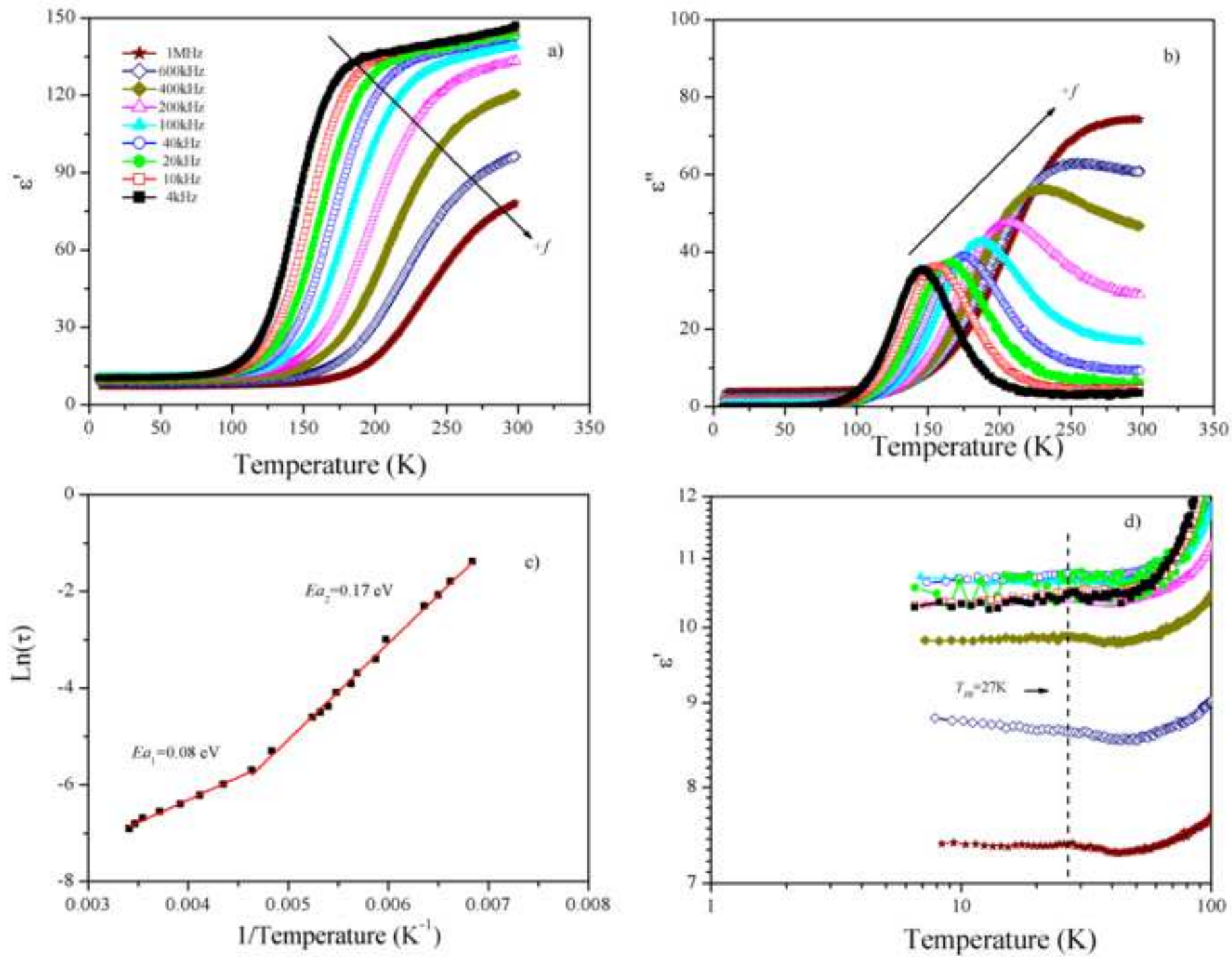